\documentclass[twocolumn,aps,epsfig,nofootinbib]{revtex4}

\usepackage{graphicx}
\usepackage{amsfonts}
\usepackage{color}

\usepackage{epstopdf}
\usepackage{latexsym}
\usepackage{amssymb}
\usepackage{amssymb}

\usepackage[center]{subfigure}

\begin{document}

 \newcommand{\bq}{\begin{equation}}
 \newcommand{\eq}{\end{equation}}
 \newcommand{\bqn}{\begin{eqnarray}}
 \newcommand{\eqn}{\end{eqnarray}}
 \newcommand{\nb}{\nonumber}
 \newcommand{\lb}{\label}
\newcommand{\PRL}{Phys. Rev. Lett.}
\newcommand{\PL}{Phys. Lett.}
\newcommand{\PR}{Phys. Rev.}
\newcommand{\CQG}{Class. Quantum Grav.}
\newcommand{\hong}[1]{\textcolor{red}{#1}}
 
 \title{Quantization of (1+1)-dimensional  Ho\v{r}ava-Lifshitz theory of gravity}

\author{Bao-Fei Li$^{a, b}$}

\author{Anzhong Wang$^{a, b}$ \footnote{The corresponding author\\ E-mail: Anzhong$\_$Wang@baylor.edu}}

\author{Yumei Wu${}^{a, b}$}

\author{Zhong Chao  Wu${}^{a}$}

\affiliation{$^{a}$ Institute  for Advanced Physics $\&$ Mathematics, Zhejiang University of Technology, Hangzhou 310032,  China\\
$^{b}$ GCAP-CASPER, Physics Department, Baylor University, Waco, TX 76798-7316, USA}

\date{\today}

\begin{abstract}

In this paper, we study  the quantization of the (1+1)-dimensional projectable Ho\v{r}ava-Lifshitz (HL) gravity, and find 
that, when only gravity is present, the system can be quantized by following  the canonical Dirac quantization, and the 
corresponding wavefunction is normalizable for some orderings of the operators. The corresponding Hamilton can also 
be written in terms of a  simple harmonic oscillator, whereby the quantization can be carried out quantum mechanically 
in the standard way.  When the HL gravity minimally couples to a scalar field, the momentum constraint is solved explicitly 
in the case where the fundamental variables are functions of time only. In this case, the coupled system can also be 
quantized by following the Dirac process, and the corresponding wavefunction is also normalizable for some particular 
orderings of the operators. The Hamilton can be also written in terms of two interacting harmonic oscillators. But, when 
the interaction is turned off,  one of the harmonic oscillators has positive energy, while the other has negative energy. A 
remarkable feature is that orderings of the operators from a classical Hamilton to a quantum mechanical one play a 
fundamental role in order for the Wheeler-DeWitt equation to have nontrivial solutions. In addition, the space-time is well 
quantized, even when it is  classically singular.

\end{abstract}


\maketitle

\section{Introduction}
\renewcommand{\theequation}{1.\arabic{equation}} \setcounter{equation}{0}

Quantum field theory   (QFT) provides a general framework for all  interactions of the nature, but   with one exception,  gravitation. However, the 
universal coupling of gravity to all forms of energy   makes it plausible that gravity should be also implemented in such a framework. 
In addition, around the singularities of the Big Bang   and black holes, space-time curvatures become so high and it is generally expected that
general relativity (GR), as a classical theory, is no longer valid, and Planck physics must be taken into account, whereby these singularities will be 
smoothed out and a physically meaningful description near these singular  points is achieved.  Moreover, once it is confirmed \cite{Planckb}, 
the recent detection of the  primordial gravitational waves by BICEP2 \cite{BICEP2}   indicates that in high energy gravity is 
quantized \cite{KW}. 

Motivated by the above considerations, quantization of gravity has been one of the main driving forces in physics in the past decades \cite{QGs},
and various  approaches have been pursued, including string/M-Theory  \cite{string} and loop quantum gravity  \cite{LQG}. However, it is fair
to say that so far a well-established quantum theory of gravity  is still absent, and many  questions  remain open.

Recently, Ho\v{r}ava \cite{Horava} proposed a theory of quantum gravity in the framework of QFTs, with the
perspective that Lorentz symmetry (LS) appears only as an emergent symmetry at low energies, but can be fundamentally
absent at high energies \footnote{Note that  the breaking of LS in the matter sector is highly restricted
by experiments, but in the gravitational sector the restrictions are much weaker \cite{Pola,LZbreaking,PS12}.}.
In Ho\v{r}ava's theory, the LS  is broken via the anisotropic scaling between time and space 
in the ultraviolet (UV), 
\bq
\lb{1.1}
t \rightarrow b^{-z} t,\;\;\; x^{i} \rightarrow b^{-1} x^{i}, \; (i = 1, 2, ..., d),
\eq
where $z$ denotes the dynamical  critical exponent. 
This is a reminiscent of Lifshitz scalars  in condensed matter physics \cite{Lifshitz}, hence the theory is often referred to as the
Ho\v{r}ava-Lifshitz (HL) gravity. For the theory to be power-counting renormalizable, the critical exponent $z$ must be $z \ge d$
\cite{Horava,Visser}, while the relativistic scaling corresponds to $z = 1$.   

With Eq.(\ref{1.1}) as the guideline, Ho\v{r}ava assumed that the breaking of the LS and thus 
the $4$-dimensional diffeomorphism invariance  is only down to the so-called foliation-preserving diffeomorphism,
\begin{equation}
\lb{1.2}
 t \to t'(t), \quad x^i \to {x'}^{i} \left(t, x^k\right),
\end{equation}
often denoted  by Diff($M, \; {\cal{F}}$). This gauge symmetry provides  a crucial 
ingredient to the construction of  the HL gravity: its action includes only  higher-dimensional spatial (but not time) derivative operators, so that the UV behavior is
dramatically improved, and in particular can be made   power-counting renormalizable. The exclusion of high-dimensional time derivative operators,
on the other hand,   prevents the ghost instability, so that the long-standing problem of unitarity is resolved  \cite{Stelle}. In the infrared (IR) the lower dimensional operators
take over, and presumably provide a healthy IR  limit.

Applying the HL theory to cosmology, various remarkable features were found \cite{reviews}.  In particular, the
higher-order spatial curvature terms can give rise to a bouncing universe \cite{CalcagniA}, and may ameliorate the flatness problem \cite{KK}. 
The anisotropic scaling provides a solution to the horizon problem and generation of scale-invariant perturbations  either with \cite{WWM}  or 
without \cite{Mukohyama:2009gg}  inflation. The scalar perturbations become adiabatic, not because of the conservation of energy as in GR \cite{MW09}, 
but because of the slow-roll condition \cite{WWM}. Similar results were also obtained for tensor perturbations  of primordial gravitational waves 
 \cite{Wang10}, while the vector perturbations are still trivial. The dark sector can have its purely geometric origins \cite{Mukohyama:2009mz,Wang}, 
and so on.

Despite  all these remarkable features, it was soon found that the original version of the HL gravity is plagued with   several undesirable issues,
including the IR instability \cite{Horava,WM} and   strong coupling \cite{SC}. 
To address  these problems, various models  were proposed \cite{reviews}. So far, several models  are  free of these problems, 
and are  consistent with the solar system tests \cite{LMWZ,Yagi} and cosmological observations \cite{InflationA,InflationB}: One is   
 the healthy extension of the HL gravity \cite{BPS}, and another is  the  nonprojectable  general covariant  theory  \cite{ZWWS}. 
 The latter has been recently embedded into string theory \cite{JK}. 

In this paper, we study another important issue of the HL gravity - the quantization. It is well-known that normally this  becomes extremely complicated and very mathematically involved in 
(3+1)-dimensions  \cite{QGs}. To bypass these technical issues,  in this paper we shall study the quantization of the HL gravity  in (1+1)-dimensional (2d) spacetimes, so the problem becomes 
tractable, and may still be able to shed lights on some basic  nature of the quantization of the theory, as various important examples of the (3+1)-dimensional (4d) gravity belong to this class, 
including spherically  symmetric black holes and the FRW universe, not to mention the string inspired models \cite{2dDilatons},  although it is also well-known that GR in 4-dimensions  
is quite different from  that in   lower dimensions \cite{SC98}. 

Specifically, the rest of the paper is organized as follows: In Sec. II we shall provide a brief review on the 2d HL gravity, from which it can be seen that, unlike the 2d GR, the 2d HL gravity is non-trivial 
even without coupling to matter.  This can be further seen from the non-trivial (classical) vacuum solutions of the theory with the projectability condition, presented in Sec. III, in which the local and global 
properties of the solutions are also studied. In Sec. IV, the quantization of the 2d HL gravity is carried out explicitly by the canonical Dirac quantization. In addition, we find that the problem can 
also be  reduced to the quantization of a simple harmonic oscillator \cite{AGSW}.
In Sec. V, we generalize these studies to the case where the HL gravity is minimally coupled to a scalar field, which shares  the same gauge symmetry as the 2d
gravitational sector. Unlike the vacuum case, we find that now the momentum constraint cannot be solved explicitly in general. Then, we restrict ourselves only to
the case   in which the fundamental variables depend only on time.
Similar to the vacuum case, now the system can also be quantized by the standard Dirac quantization. 
The corresponding Hamilton can be also written in two interacting harmonic oscillators. When the interaction  vanishes,  one of the two  
 oscillators has positive energy and the other has negative energy.  
 The paper is ended in Sec. VI, in which we derive our main conclusions. A remarkable feature is that orderings of the operators from a classical Hamilton to a quantum mechanical one play a 
fundamental role in order for the Wheeler-DeWitt equation $\hat{H}\left|\Psi\right>  = 0$ to have nontrivial solutions.
 In addition,  space-times can be still well quantized, although
they are   classically singular   [cf. Fig.\ref{fig1}]. This is true not only for the vacuum case, but also for the case coupled with the scalar 
field [cf. Fig.\ref{fig4}]. 
 
Note that the quantization of the 2d HL gravity was studied recently in \cite{AGSW}, and showed that it is equivalent to the 2d causal dynamical triangulations (CDT) when the projectability condition
is imposed.  In addition, the 3d projectable HL gravity was also studied  numerically in terms of CDT \cite{ACCHKZ}, and  found  evidence for the consistency of the quantum  
phases of solutions to the equations of motion of classical HL gravity.  Benedetti and  Guarnieri, on the other hand,
 studied one-loop renormalization in a toy model of the HL gravity,  that is,  the conformal reduction of the 
$z=2$ projectable HL theory \cite{BG}. They  found that the would-be asymptotic freedom associated to the running Newton's constant is exactly balanced by the strong coupling of the 
scalar mode as the Weyl invariant limit is approached. Then, they concluded that  in such model the UV limit is singular at one loop order, and   argued that a similar phenomenon should be
 expected in the full theory, even in higher dimensions. Loop corrections and renormalization  group flows were also studied in some particular models of the HL gravity \cite{PS12,OR}, and different
 conclusions were obtained for different models.  

\section{Horava-Lifshitz Theory of Gravity in (1+1)-Dimensions}
\renewcommand{\theequation}{2.\arabic{equation}} \setcounter{equation}{0}

The Einstein's theory of gravity in (1+1)-dimensional spacetimes is trivial, as the   
Riemann and Ricci tensors ${\cal{R}}_{\mu\nu\beta\gamma}$ and ${\cal{R}}_{\mu\nu}$    are uniquely determined
by the Ricci scalar ${\cal{R}}$ via the relations \cite{SC98},
\bqn
\lb{2.1}
{\cal{R}}_{\mu\nu\beta\gamma} &=& \frac{1}{2}\left(g_{\mu\beta}g_{\nu\gamma} - g_{\mu\gamma}g_{\nu\beta}\right){\cal{R}},\nb\\
{\cal{R}}_{\mu\nu} &=& \frac{1}{2}g_{\mu\nu}{\cal{R}}, 
\eqn
where the Greek letters run from 0 to 1. Then, the Einstein  tensor $E_{\mu\nu} [= {\cal{R}}_{\mu\nu} -  \frac{1}{2}g_{\mu\nu}{\cal{R}}]$
always vanishes, and the Einstein-Hilbert action \footnote{ In 2d spacetimes, the integral $\int{d^2x \sqrt{{}^{(2)}g}\;\; {\cal{R}}}$ always gives a 
boundary term. So, normally one does not consider it. This can also be seen from the field equations (\ref{2.3}).}
\bq
\lb{2.2}
S_{EH} = \zeta^2\int{d^2x\sqrt{{}^{(2)}g} \left({\cal{R}} - 2\Lambda + \zeta^{-2} {\cal{L}}_{M}\right)}, 
\eq
leads to a set of non-dynamical field equations,  in which the metric $g_{\mu\nu}$ is directly related to the energy-momentum tensor $T_{\mu\nu}$ via the relation,
\bq
\lb{2.3}
\Lambda g_{\mu\nu} = 8\pi G T_{\mu\nu},
\eq
where $\zeta^2 = 1/(16\pi G)$ \footnote{It should be noted that, unlike in the 4-dimensional case, now  $\zeta$ is dimensionless (so is $G$).}.
Therefore, in order to have a non-trivial theory of gravity in 2-dimensions (2d), extra degrees are often introduced, such as a dilaton \cite{2dDilatons} or a Liouville field \cite{LG}.

However, this is not the case for the HL gravity \cite{Horava}, as the latter has a different symmetry, the foliation-preserving diffeomorphisms (\ref{1.2}).
Then, the general gravitational action takes the form,
\bq
\lb{2.5}
S_{HL}=\zeta^2 \int {dt dx   N\sqrt{g}\left({\cal{L}}_{K} - {\cal{L}}_{V}\right)},
\eq
where   $N$ denotes  the lapse function in the Arnowitt-Deser-Misner (ADM) decompositions \cite{ADM}, and
 $g \equiv {\mbox{det}}(g_{ij})$, here $g_{ij}$ is the spatial metric defined on the leaves $t=$ Constant.   ${\cal{L}}_{K}$ is the kinetic part of the action, given by
\bq
\lb{2.6}
{\cal{L}}_{K} = K_{ij}K^{ij} - \lambda K^2,
\eq
where $\lambda$ is a dimensionless constant, and $K_{ij}$ denotes the extrinsic curvature tensor of the leaves $t=$ Constant, given by 
\bq
\lb{2.7}
K_{ij}=\frac{1}{2N}\left(-\dot g_{ij}+\nabla_iN_j+\nabla_jN_i\right),
\eq
and   $K \equiv g^{ij}K_{ij}$. Here $\dot{g}_{ij} \equiv \partial{g}_{ij}/\partial t$, $\nabla_i$ denotes the covariant derivative of the metric $g_{ij}$, and $N^i$ the shift vector.
In the (1+1)-dimensional case, since there is only one spatial dimension, we have $i, j = 1$, and 
\bqn
\lb{2.8}
K&=& g^{11}K_{11} = -\frac{1}{N}\left(\frac{\dot\gamma}{\gamma}-\frac{N_1'}{\gamma^2}+\frac{N_1\gamma'}{\gamma^3}\right),
\eqn
where $\gamma \equiv \sqrt{g_{11}},\;   \gamma' \equiv \partial\gamma/\partial x$, etc. 

On the other hand,  ${\cal{L}}_{V}$ denotes the potential part of the action, and is made of $R, \; \nabla_i$ and $a_i$, 
that is, 
\bq
\lb{2.9}
{\cal{L}}_{V} = {\cal{L}}_{V}\left(R, \; \nabla_i, \; a_i\right),
\eq
where $a_i \equiv N_{,i}/N$ and $R$ denotes the Ricci scalar of the  leaves $t=$ Constant, which identically vanishes in one-dimension, i.e., $R = 0$. Power-counting renormalizibility 
condition requires that ${\cal{L}}_{V}$ should contain spatial operators with the highest dimensions that are not less than $2z$, where $z \ge d$ \cite{Horava,Visser}, and $d$
denotes the number of the spatial dimensions. Taking the  minimal requirement, that is, $z =d$, we find that in the current case ($d = 1$) we have 
\bq
\lb{2.10}
{\cal{L}}_{V} = 2\Lambda - \beta a_i a^i,  
\eq
where $\Lambda$ denotes the cosmological constant, and $\beta$ is another dimensionless coupling constant. Collecting all the above together, we find that the gravitational
action of the HL gravity in $(1+1)$-dimensional spacetimes can be cast in the form, 
\bq
\lb{2.11}
S_{HL}=\zeta^2 \int {dt dx   N\sqrt{g}\left[(1-\lambda)K^2-2{\Lambda} + \beta a_ia^i\right]}.
\eq

\section{Classical Solutions of the 2d HL Gravity with the projectable condition}
\renewcommand{\theequation}{3.\arabic{equation}} \setcounter{equation}{0}

Assuming the projectabilility condition, we have \cite{Horava}
\bq
\lb{2.12}
N = N(t), 
\eq
from which we immediately find $a_i = 0$. In the rest of this section, we shall assume this condition.  Then, the variations of the action $S_{HL}$ with respect to $N$ and  $N_1$ yield the 
Hamiltonian and momentum constraints,  and are given, respectively,  by
\bqn
\lb{2.13a}
\int dx \gamma(K^2+4\tilde{\Lambda})=0,  \\
\lb{2.13b}
K' =0,
\eqn
where $\tilde{\Lambda} \equiv {\Lambda}/{[2(1-\lambda)]}$. The variation of the action $S_{HL}$ with respect to $\gamma$, on the other hand, yields the dynamical equation, 
\bqn
\label{2.14}
&& \dot K +\frac{1}{2} N(K^2-4\tilde{\Lambda}) +\frac{K \dot\gamma}{\gamma} - \frac{2KN_1' }{\gamma^2}\nb \\
&& ~~~~ +\left(\frac{N_1K}{\gamma^2}\right)' + \frac{3KN_1\gamma'}{\gamma^3} =0.
\eqn

Using the gauge freedom of Eq.(\ref{1.2}), without loss of the generality, we can always set
\bq
\lb{2.14a}
N = 1, \;\;\; N_1 = 0,
\eq
so that the 2d metric takes the form,  
\bq
\label{2.15}
ds^2=-dt^2+\gamma^2(t, x) dx^2.
\eq
 It should be noted that Eq.(\ref{2.14a}) uniquely fixes the gauge only up to 
 \bq
 \lb{2.16}
 t' = t + t_0, \;\;\; x' = \zeta(x),
 \eq
where $t_0$ is a constant, and $\zeta(x)$ is an arbitrary function of $x$ only.

With the above gauge choice, Eq.(\ref{2.14}) reduces to   
\bq
\lb{2.17}
K^2-2\dot K+ 4\tilde{\Lambda}=0.
\eq 
On the other hand, from the momentum constraint (\ref{2.13b}) we can see that $K$ is independent of $x$, so  the  Hamiltonian constraint Eq.(\ref{2.13a}) reduces to, 
\bq
\lb{2.13c}
(K^2+4\tilde{\Lambda}) \int dx{\gamma(t,x)}=0. 
\eq
Therefore, there exist two possibilities,
\bq
\lb{2.13ca}
i) \; K^2+4\tilde{\Lambda} = 0,\;\;\;
ii)\; \int dx{\gamma(t,x)}=0.
\eq
In the following, we consider them separately.

\subsection{$K^2+4\tilde{\Lambda}=0$}

In this case, the extrinsic curvature $K$ is just a constant given by 
\bq
K=\pm 2 \sqrt{-\tilde{\Lambda}}, 
\eq
which makes sense only when $\tilde{\Lambda}<0$. From Eq.(\ref{h}), we can find
\bq
\gamma=e^{\pm 2\sqrt{-\tilde{\Lambda}}t+F(x)},
\eq
here $F(x)$ is an  arbitrary function of $x$ only. Using the gauge residual (\ref{2.16}), we can always set $F(x)=0$, so the metric reduces to,
\bqn
\lb{2.17aa}
ds^2 = -dt^2+e^{4\sqrt{-\tilde{\Lambda}}t}dx^2.
\eqn
This is nothing but the de Sitter spacetime.

\subsection{$\int dx{\gamma(t,x)}=0$}

In this case, we can see that $\gamma(t,x)$ has to be an odd function of $x$, i.e., 
$\gamma(t,x) = - \gamma(t, -x)$. Then, from Eq.(\ref{2.17}) we find that
\bq
\label{g}
\frac{dK}{K^2+4\tilde{\Lambda}}=\frac{1}{2} dt. 
\eq
Since $K$ is independent of $x$, we find
\bq
\label{h}
 \frac{\dot\gamma}{\gamma} = - K(t).
\eq
To solve the above  equations under the constraint $\int dx{\gamma(t,x)}=0$, it is found convenient to consider the cases
$\tilde{\Lambda}>0$, $\tilde{\Lambda}<0$, and $\tilde{\Lambda}=0$, separately.

\subsubsection{ $\tilde{\Lambda}>0$}

\begin{figure}[tbp]
\centering
\includegraphics[width=8cm]{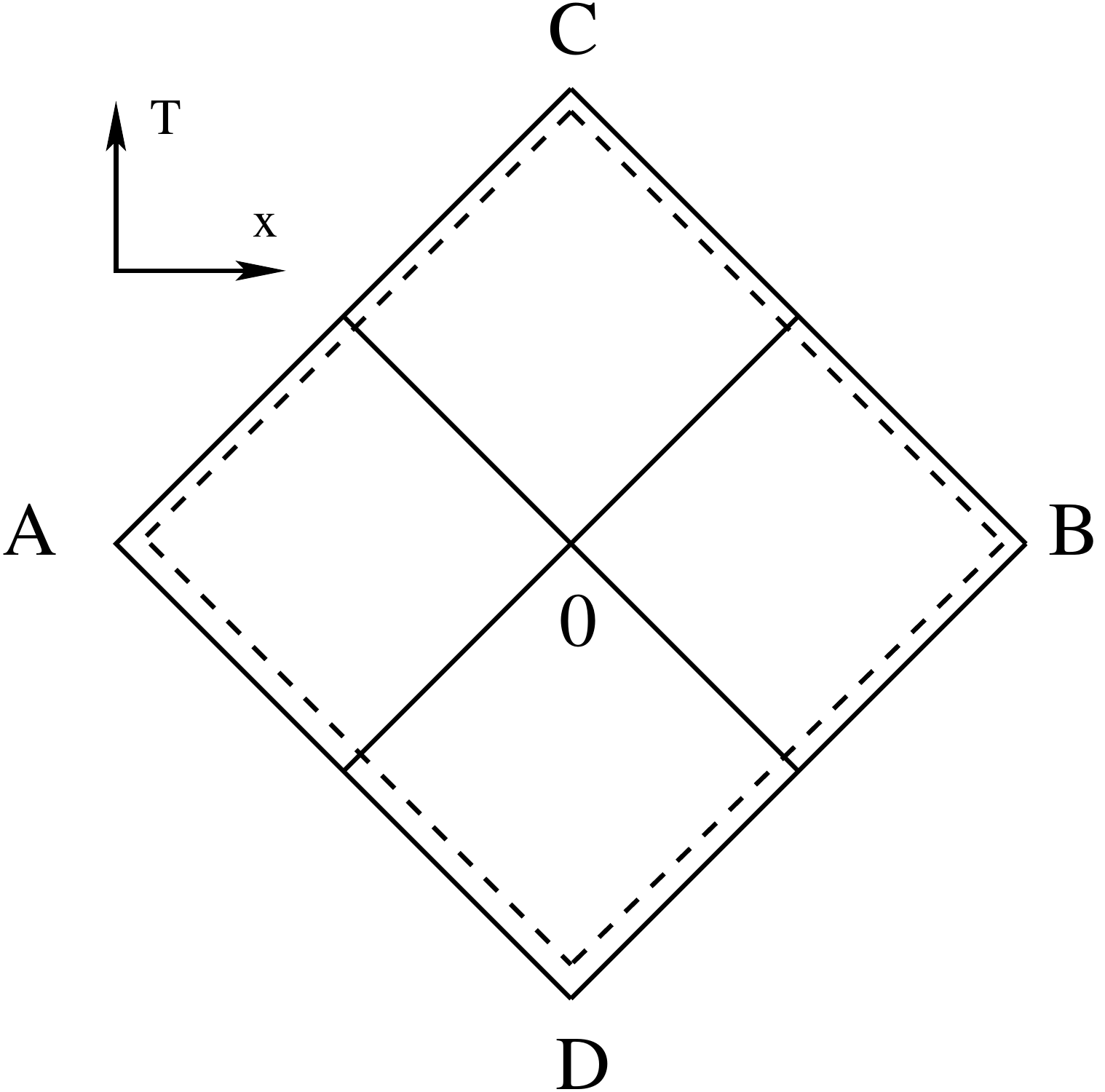}
\caption{The Penrose diagram for the solution (\ref{2.18}), in which the space-time is singular at both past and further null infinities ($T = \pm \infty$),
denoted by the lines $\overline{AC}, \; \overline{AD},\; \overline{BC}$ and $\overline{BD}$.}
 \label{fig1}
\end{figure}

Straightforward integration of Eq. (\ref{g}) gives us 
\bq
\lb{2.15a}
K= \beta \tan\left[\frac{\beta}{2}(t-t_0)\right], 
\eq
where $\beta \equiv \sqrt{4|\tilde{\Lambda}|}$.
Then, from Eq.(\ref{h}) we find, 
\bq
\lb{2.16a}
\gamma=\cos^2\left(\frac{\beta (t-t_0)}{2}\right){\hat{\gamma}(x)}.
\eq
To satisfy the Hamiltonian constraint, $\hat{\gamma}(x)$ must be an odd function of $x$, so that
\bq
\lb{2.16aa}
\int_{-L_{\infty}}^{L_{\infty}}{\hat{\gamma}(x)dx} = 0,
\eq
where $x = \pm L_{\infty}$ denote the boundaries of the spacetime in the spatial direction, which can be taken to infinity.   With this in mind, we can introduce a new coordinate $x'$ by $
dx' = \hat{\gamma}(x) dx$,   
so the metric takes the form,
\bqn
\lb{2.17a}
ds^2 = -dt^2+cos^4\left(\frac{\beta t}{2}\right){dx'}^2.
\eqn
Note that in writing the above expression, we had set $t_0 = 0$ by using another gauge freedom given in Eq.(\ref{2.16}). 
Setting 
\bq
\lb{2.17b}
T =  \frac{2}{\beta}\tan\left(\frac{\beta t}{2}\right),
\eq
the above metric can be cast in the conformally-flat form, 
\bq
\lb{2.18}
ds^2=\left(1+\frac{\beta^2}{4}T^2\right)^{-2} \left(-dT^2+ {dx'}^2\right),
\eq
for which we have 
\bq
\lb{2.19}
K=  \frac{\beta^2}{2}T.
\eq
That is, the space-time is singular at $ T = \pm \infty$. This is a real space-time singularity in the HL gravity \cite{CW10}, since
it is a scalar one and cannot be removed by any coordinate transformations allowed by the symmetry of the theory. The corresponding 
Penrose diagram is given by Fig. \ref{fig1}. 

\subsubsection{$\tilde{\Lambda}<0$}

In this case,  Eq.(\ref{g})  has the solution
\bq
\lb{2.20}
K=\cases{-\beta \tanh\left[\frac{\beta}{2}(t-t_0)\right], & $ |K| < \beta$,\cr
-\beta \coth\left[\frac{\beta}{2}(t-t_0)\right], & $ |K| > \beta$.\cr}
\eq
 In the following, let us consider the two cases separately, as they will have different  properties.

{\bf Case a) $ |K| < \beta$:} Then, from Eq.(\ref{h}) we find that
\bq
\lb{2.21}
\gamma=\cosh^2\left[\frac{\beta}{2}(t-t_0)\right] \hat{\gamma}(x).
\eq
Again, using the gauge residual (\ref{2.16}), without loss of the generality, we can always set $t_0 = 0$ and $dx' = \hat{\gamma}(x) dx$, so the metric finally takes the form,  
\bqn
\lb{2.22}
ds^2 = -dt^2+\cosh^4\left(\frac{\beta t}{2}\right) {dx}^2.
\eqn
Note that we dropped the prime from $x$ in writing down the above expression. Then, 
  we can see that the metric is singular at $t = \pm \infty$. However, Eq.(\ref{2.20}) shows that $K$ is finite at these two limit. In addition, 
the corresponding 2d Ricci scalar ${\cal{R}}$ is given by
\bq
\lb{2.23}
{\cal{R}} = \beta^2 \frac{\cosh(\beta t)}{\cosh^2\left(\frac{\beta t}{2}\right)},
\eq
which is also finite as $t \rightarrow \pm \infty$. To further study the properties of these singularities, let us consider the tidal forces experienced by
a free-falling observer, whose trajectory is given by the timelike geodesics, satisfying the Euler-Lagrange equation,
\bq
\lb{2.24}
\frac{\partial {\cal{L}}_p}{\partial x^\mu} - \frac{d}{d\tau}\left(\frac{\partial {\cal{L}}_p}{\partial \dot{x}^\mu}\right) = 0,
\eq
where $\tau$ denotes the affine parameter along the geodesics, and 
\bq
\lb{2.25}
{\cal{L}}_p \equiv \left(\frac{ds}{d\tau}\right)^2 =  - \dot{t}^2 + \cosh^4\left(\frac{\beta t}{2}\right)\dot{x}^2,
\eq
but now with $\dot{t} \equiv dt/d\tau$, etc. For timelike geodesics we have ${\cal{L}}_p = -1$. Since the metric (\ref{2.22}) does not depend on $x$ explicitly, 
Eq.(\ref{2.24}) yields the conservation law of momentum, 
\bq
\lb{2.26}
 2\cosh^4\left(\frac{\beta t}{2}\right)\dot{x} = p,
\eq
where $p$ denotes the momentum of the observer. Inserting the above expression into Eq.(\ref{2.25}), we find that
\bq
\lb{2.27}
\dot{t} = \pm \frac{\sqrt{4\cosh^4\left(\frac{\beta t}{2}\right) + {p^2}}} {2\cosh^2\left(\frac{\beta t}{2}\right)},
\eq
where ``+" (``-") corresponds to the observer moving along the positive (negative) direction of the $x$-axis. 
Setting $e_{(0)}^{\mu} = dx^{\mu}/d\tau$, we can construct another space-like unit vector, $e_{(1)}^{\mu}$ as
\bq
\lb{2.28}
e_{(1)}^{\mu} = \left(\pm \frac{p}{2\cosh^2\left(\frac{\beta t}{2}\right)}, \frac{\sqrt{4\cosh^4\left(\frac{\beta t}{2}\right) + {p^2}}}{2\cosh^4\left(\frac{\beta t}{2}\right)}\right),
\eq
which is orthogonal to $e_{(0)}^{\mu}$, and parallelly transported along the time-like geodesics,  
\bq
\lb{2.29}
g_{\mu\nu} e^{\mu}_{(a)}  e^{\nu}_{(b)} = \eta_{ab}, \;\;\; e^{\mu}_{(1); \nu} e^{\nu}_{(0)} = 0, 
\eq
where $\eta_{ab} = {\mbox{diag.}}(-1, 1)$, and a semicolon ``;" denotes the covariant derivative with respect to the two-dimensional metric $g_{\mu\nu}$. Projecting the two-dimensional Ricci 
tensor onto the above orthogonal frame, we find that
\bqn
\lb{2.30}
R_{(0)(0)} &=& - R_{(1)(1)} =  - \frac{\beta^2 \cosh(\beta t)}{2\cosh^2\left(\frac{\beta t}{2}\right) },\nb\\
R_{(0)(1)} &=& 0,
\eqn
which are all finite as $t \rightarrow \pm \infty$. Therefore, the singularities at $t =  \pm \infty$ must be coordinate ones. In fact, they represent the boundaries of the
space-time. To see this, let us consider the proper time that the observer needs to travel from a given time $t_0$ to $t = \infty$, which is given by
\bq
\lb{2.31}
\Delta \tau = \int_{t_0}^{\infty}{\frac {2\cosh^2\left(\frac{\beta t}{2}\right)} {\sqrt{4\cosh^4\left(\frac{\beta t}{2}\right) + {p^2}}}} = \infty,
\eq
for any finite $t_0$. That is, starting at any given finite moment, $t_0$, the observer always needs to spend infinite proper time to reach at the time $t = \infty$. 
In other words, $t = \infty$ indeed represents the future timelike infinity of the space-time. Similarly, one can see that $t = - \infty$ represents the past timelike infinity.

\begin{figure}[tbp]
\centering
\includegraphics[width=8cm]{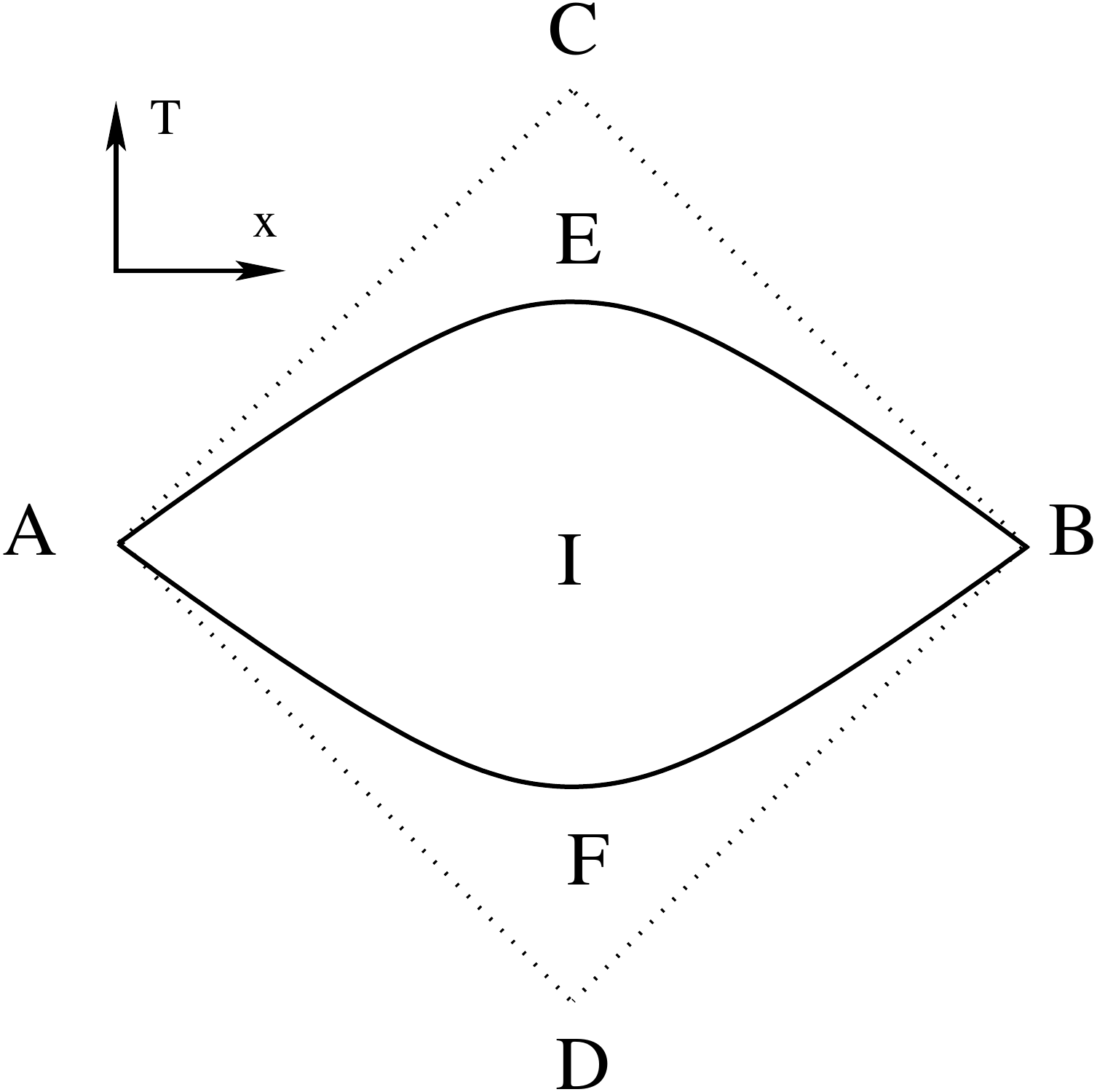}
\caption{The Penrose diagram for the solution (\ref{2.22}), in which the singularities at $t = \pm \infty$, denoted by the curves $\widehat{AEB}$ and $\widehat{AFB}$,
 are coordinate ones, and represent the physical boundaries of the space-time.}
 \label{fig2}
\end{figure}

To study its global structure,  let us first introduce the new timelike coordinate $T$ via the relation, 
\bqn
 \lb{2.32}
T=\frac{2}{\beta}\tanh\left(\frac{\beta t}{2}\right),
\eqn
we find that the metric takes the form, 
\bq
\lb{2.33}
ds^2={\left(1-\frac{\beta^2}{4}T^2\right)^{-2}}(-dT^2+dx^2),\; \left(|T| \le 2/\beta\right).
\eq
It is interesting to note that the above metric is singular at $T = \pm 2/\beta$. But, as shown above, this corresponds to coordinate singularities. In fact, they are 
the space-time boundaries, and any observer will need infinite proper time to reach them starting from any finite time. The corresponding Penrose diagram is given by 
Fig. \ref{fig2}. 

Finally, we note that the similarity of the metric (\ref{2.22}) with the $dS_2$ metric, 
\bq
\lb{2.33a}
ds^2_{dS_2} = - dt^2 + \cosh^2(\beta t) d\chi^2,
\eq
where $0 \le \chi \le \pi$ with the hypersurfaces $\chi = 0$ and $\chi  = \pi$  identified, so the whole space-time has a $R^1\times S^1$ topology.  The space-time is complete
in these coordinates. This can be seen clearly  by embedding Eq.(\ref{2.33a})  into a 3-dimensional Minkowski space-time $ds_3^2 = - dv^2 + dw^2 + dX^2$ with \cite{HE73}, 
\bqn
\lb{2.33c}
v &=& \frac{1}{\beta}\sinh(\beta t),\;\;\;
w =  \frac{1}{\beta}\cosh(\beta t) \cos\left(\frac{\chi }{\beta}\right),\nb\\
X &=&  \frac{1}{\beta}\cosh(\beta t) \sin\left(\frac{\chi }{\beta}\right),
\eqn
which is a  hyperboloid,
\bq
\lb{2.33b}
- v^2 + w^2 + X^2 = \beta^{-2},
\eq
in the 3-dimensional Minkowski space-time. 
The two metrics (\ref{2.22}) and (\ref{2.33a}) becomes asymptotically identified when $|t| \gg \beta^{-1}$, provided that the coordinate $\chi$ is unrolled  to $- \infty < \chi < \infty$.

{\bf Case b) $ |K| > \beta$:}    In this case, following what was done in the last case, it can be shown that 
\bqn
\lb{2.34}
K=-\beta \coth\left(\frac{\beta t}{2}\right), \;\;\;
 \gamma =  \sinh^2\left(\frac{\beta t}{2}\right)\hat{\gamma}(x), 
\eqn
and the corresponding  line element takes the form, 
\bqn
\lb{2.35}
ds^2&=&-dt^2+ \sinh^4\left(\frac{\beta }{2}t\right){dx'}^2.
\eqn
Similar to the last case, the metric is singular at $t = \pm \infty$. However, these are   coordinate ones, as in the last case. In fact,
following what we did there, we find that the following forms a freely-falling frame,  
\bqn
\lb{2.36}
e_{(0)}^{\mu}&=& \left(\pm \sqrt{1+\frac{p^2}{4\sinh^4\left(\frac{\beta t}{2}\right)}},\frac{p}{2\sinh^4\left(\frac{\beta t}{2}\right)}\right),\nb\\
e_{(1)}^{\mu}&=&\left(\pm \frac{p}{2\sinh^2\left(\frac{\beta t}{2}\right)}, \frac{\sqrt{1+\frac{p^2}{4\sinh^{4}\left(\frac{\beta t}{2}\right)}}}{\sinh^2\left(\frac{\beta t}{2}\right)}\right),
\eqn
for which we have 
\bqn
\lb{2.37}
R_{(0)(0)}&=& - R_{(1)(1)} = -\frac{1}{2}\beta^2\cosh(\beta t)\cosh^{-2}\left(\frac{\beta t}{2}\right), \nb\\
R_{(1)(0)}&=&0.
\eqn
It is clear that all of these components, representing the tidal forces exerted on the observer, are finite. From Eq.(\ref{2.36}) one can also show that 
\bq
\lb{2.31a}
\Delta \tau = \int_{t_0}^{\infty}{\frac {2\sinh^2\left(\frac{\beta t}{2}\right)} {\sqrt{4\sinh^4\left(\frac{\beta t}{2}\right) + {p^2}}}} = \infty,
\eq
for any finite $t_0$. That is, starting at any given finite moment, $t_0$, the observer will  reach  $t = \infty$ after spending infinite proper time, i.e., 
 $t = \infty$  represents the   space-time boundary. Similarly, one can show that $t = - \infty$ represents the past timelike infinity.

However, in contrast to the last case, the space-time now becomes singular at $t = 0$. This singularity is a scalar singularity, as one can see from Eq.(\ref{2.34})
and the expression for the 2-dimensional  Ricci scalar,
\bq
\lb{2.36a}
{\cal{R}} = \beta^2\left[1 + \coth^2\left(\frac{\beta }{2}t\right)\right].
\eq

\begin{figure}[tbp]
\centering
\includegraphics[width=8cm]{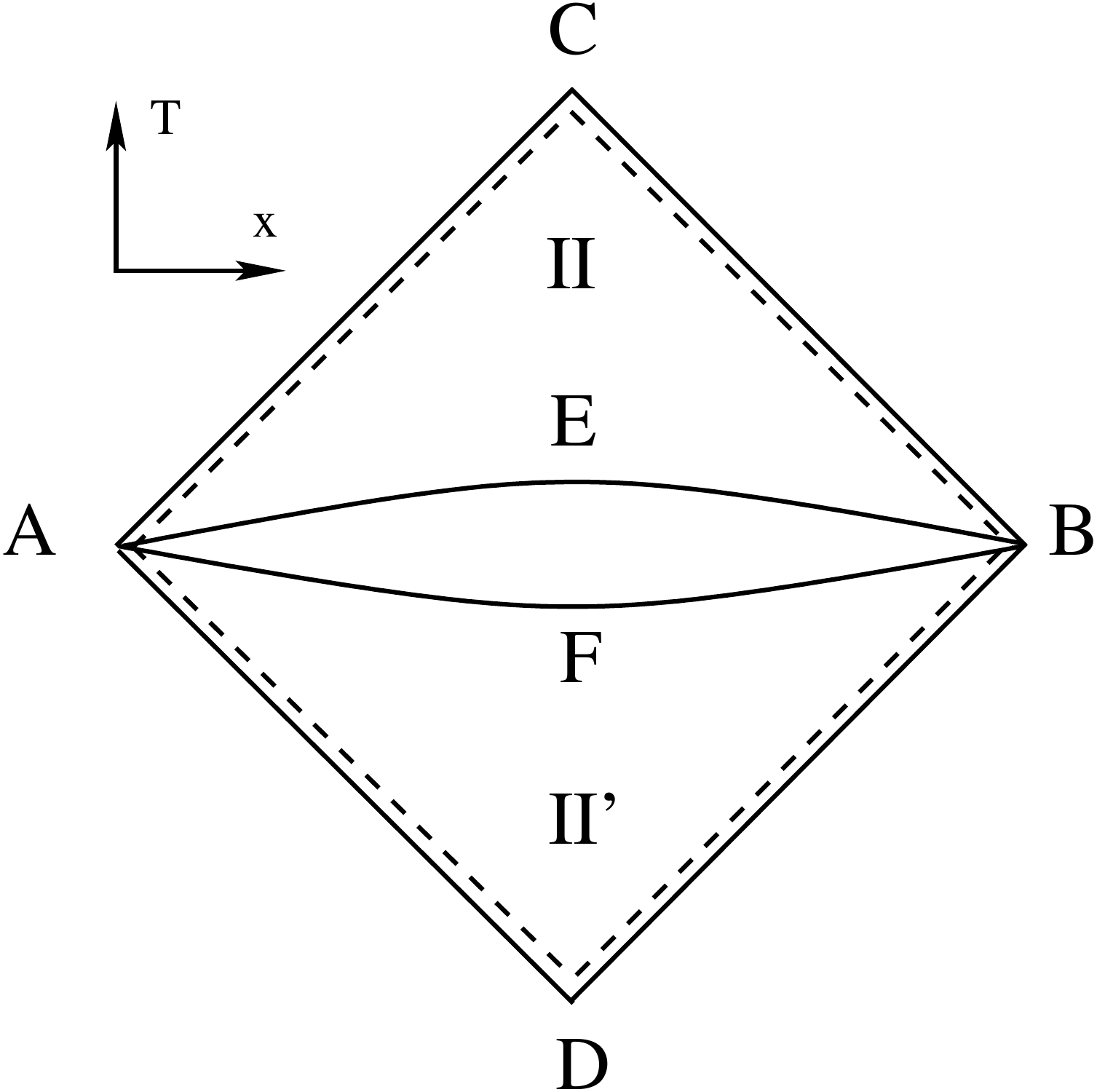}
\caption{The Penrose diagram for the solution (\ref{2.35}), in which the space-time is singular at both past and further null infinities ($T = \pm \infty$ or
$t = 0$), denoted by the lines $\overline{AC}, \; \overline{AD},\; \overline{BC}$ and $\overline{BD}$. The curved lines, $\widehat{AEB}$ and $\widehat{AFB}$, 
are free of space-time singularities, and represent the physical boundaries   of the space-time.  }
 \label{fig3}
\end{figure}

To study its global properties, we first introduce the new coordinate $T$ via the relation 
\bqn
\lb{2.37a}
T= -\frac{2}{\beta} \coth\left(\frac{\beta }{2}t\right),
\eqn
which maps $t \in (-\infty, 0)$ into the region $T \in (2/\beta, \infty)$, and $t \in (0, \infty)$ into the region $T \in (-\infty, - 2/\beta)$. In particular, the time
$t = 0^{\pm}$ are mapped to $T = \mp \infty$, and $t = \pm \infty$ to $T = \mp 2/\beta$. 
In terms of $T$, we find that
\bq
\lb{2.38}
ds^2=\left[1-\frac{\beta^2}{4}T^2\right]^{-2}\left(-dT^2+{dx'}^2\right), \; \left(|T| \ge 2/\beta\right).
\eq
The corresponding Penrose diagram is given by Fig. \ref{fig3}, from which we can see that the nature of the singularity at $t = 0$ is null. 

It is remarkable to note that the metrics (\ref{2.33}) and (\ref{2.38}) take the same form, but with different covering ranges. In Eq.(\ref{2.33}) we have $|T| \in (0, 2/\beta)$, while in 
Eq.(\ref{2.38}) we have $|T| \in (2/\beta, \infty)$. The metrics are singular at $|T| =  2/\beta$, which represent the boundaries of the spacetimes, represented, respectively, by
Eqs.(\ref{2.33}) and (\ref{2.38}). 

\subsubsection{ $\tilde{\Lambda}=0$}
 
Following what we have done in the above, it can be shown that
\bq
K=-\frac{2}{t}, \;\;\; \;\;\; \;\;\;\;
 \gamma = t^2\hat{\gamma}(x), 
\eq
and the line element takes the form
\bq
ds^2=-dt^2+t^4{dx'}^2.
\eq
Setting 
\bq
T=1/t,
\eq
 then in the new coordinates we find that the metric takes the form
\bq
ds^2=\frac{1}{T^4}\big(-dT^2+{dx'}^2\big),
\eq 
for which we have 
\bq
K=-2T.
\eq
That is, the space-time is singular at $T=\pm \infty$, and the corresponding Penrose diagram is similar to that given in Fig.\ref{fig1}.

\section{Quantization of 2d HL gravity}
\renewcommand{\theequation}{4.\arabic{equation}} \setcounter{equation}{0}

In the projectable HL gravity, the action (\ref{2.11})  reduces to   
\bq
\lb{4.1}
S_{HL}=\zeta^2 \int {dt dx   N\gamma \left[(1-\lambda)K^2-2{\Lambda}\right]},
\eq
where  $K$ is given by Eq.(\ref{2.8}). In the following, we'll quantize the field by following Dirac's approach.

\subsection{Hamiltonian Formulation  and Dirac Quantization}

Starting from the action Eq.(\ref{4.1}), if we treat $\gamma$ as a dynamical variable, its canonical momentum is found to be 
\bq
\label {d}
\pi\equiv \frac{\partial \cal{L}}{\partial \dot{\gamma}}=2\zeta^2(\lambda-1)K.
\eq
After the Legendre transformation, the corresponding canonical Hamilton is given by
\bq
\lb{4.19aa}
H_c(t)=\int{dx \Big(N \mathcal{H}(x)+N_1(x) \mathcal{H}_1(x) \Big)},
\eq
here the time variable is suppressed. With the projectability condition, the momentum constraint is local while the Hamiltonian constraint is global, that is,
\bqn
\lb{4.19a}
&& \mathcal{H}_1= -\frac{\pi'}{\gamma}\approx 0,\\
\lb{4.19b}
&& \int{dx \mathcal{H}(x)}=\int{dx \Big(\frac{\pi^2\gamma}{4\zeta^2(1-\lambda)}+2\Lambda \zeta^2 \gamma \Big)} \nb\\
&& ~~~~~~~~~~~~~ \approx 0.
\eqn
Straightforward calculations give us their Poisson brackets,
\bqn
\lb{4.20a}
\{\mathcal{H}(x),\mathcal{H}(x')\}&=&0,\nb\\
\{\mathcal{H}(x),\mathcal{H}_1(x')\}&=&\frac{\mathcal{H}(x')}{\gamma^2(x')}\delta_{x'}(x-x')\nb\\
&& +\frac{\pi \mathcal{H}_1\delta(x-x')}{\zeta^2(1-\lambda)}\approx 0, \nb\\
\{\mathcal{H}_1(x),\mathcal{H}_1(x')\}&=&\frac{2\mathcal{H}_1(x')\delta_{x'}(x-x')}{\gamma^2(x')}\nb\\
&& -\frac{2\gamma' \mathcal{H}_1}{\gamma^3}\delta(x-x') + \frac{\mathcal{H}'_1}{\gamma^2}\delta(x-x')\nb\\
&\approx& 0.
\eqn
Therefore,  we've got all the constraints and the physical degrees of freedom of the theory per space-time point  ($\mathcal{N}$)  are 
\bqn
\mathcal{N}&=&\frac{1}{2}\big({\mbox{dim}}\mathcal{P}-2\mathcal{N}_1-\mathcal{N}_2\big)\nb\\
&=&\frac{1}{2}\big(4-2*2-0\big)=0.
\eqn
Here ${\mbox{dim}}\mathcal{P}$ means the dimension of the phase space, $\mathcal{N}_1$($\mathcal{N}_2$) denotes the number of first-class (second-class) constraints.
 Meanwhile, the local momentum constraint indicates that $\pi$ is a function of time only, i.e., 
\bq
\lb{4.20b}
\pi(x, t)=\pi(t).
\eq
Note also that the canonical momentum $\pi(t)$ is  invariant under the gauge transformation, as can been seen from the expression, 
\bq
\left\{\pi(x),\int{dx' \xi(x')\mathcal{H}_1(x')}\right\}=\frac{\xi(x)\mathcal{H}_1(x)}{\gamma(x)},
\eq
which vanishes on the constraint surface. For completeness, we also give the variation of $\gamma$ under the spatial diffeomorphism,
\bq
\lb{5.3b}
\left\{\gamma(x),\int{dx' \xi(x')\mathcal{H}_1(x')}\right\}=\left(\frac{\xi}{\gamma}\right)'.  
\eq

Since the momentum $\pi$ is only a function of time, we can obtain an equivalent constraint by integrating Eq.(\ref{4.19b}) directly, and then we have
\bq
\lb{5.3ba}
H(\pi, L)=\frac{\pi^2 L}{4\zeta^2(1-\lambda)}+2\Lambda \zeta^2 L\approx 0,
\eq
with 
\bq
\lb{4.5}
L(t)=\int dx \gamma(t,x),
\eq 
which  is gauge-invariant owing to Eq.(\ref{5.3b}). It's worth noting that $\pi(t)$ can be regarded as conjugate momentum to the invariant length $L(t)$. Starting from the basic relation
\bqn
\{\gamma(x),\pi(y)\}=\delta(x-y),
\eqn
then integrating both sides with respect to x and y, since $\pi$ is independent of spatial coordinate y, we directly get 
\bq
\lb{5.3bb}
\{L(t), \pi(t)\}=1.
\eq
Now following Dirac's approach, by promoting Eq.(\ref{5.3bb})  to the commutation relation $[\hat{L},\hat{ \pi}]=i$, we get the Wheeler-DeWitt equation in the coordinate representation,
\bq
\lb{5.3bc}
\hat{H}\Psi=0.
\eq
However, there is ordering ambiguity arising from the term $L\pi^2$ in Eq.(\ref{5.3ba}) \cite{AGSW}. In the following we consider each of the possible orderings, separately.

 \subsubsection{$:\pi^2L:  \; = \hat{L}{\hat{\pi}}^2$}

In this case, the Hamiltonian constraint reads
\bq
L \left(\frac{\partial^2}{ \partial L^2}-\epsilon_{\tilde{\Lambda}}\mu^2\right)\Psi=0,
\eq
 where $\mu\equiv 4\zeta^2 |1-\lambda |\sqrt{|\tilde{\Lambda}|}$, and $\epsilon_{\tilde{\Lambda}}$ is a sign function which is one for $\tilde{\Lambda}>0$, zero for  $\tilde{\Lambda}=0$ and negative one 
 for $\tilde{\Lambda}<0$.
 For $\tilde{\Lambda}>0$,
the general solution is  
\bq 
\Psi(L, t) = C_1 e^{ \mu L} + C_2 e^{-\mu L}.
\eq
It can be shown that this solution is not normalizable  even with   $C_1 = 0$ with  respect to the measure $L^{-1}dL$ in the interval $(0,+\infty)$.  For $\tilde{\Lambda}=0$, we have
\bq
\Psi(L, t)=A_1L+A_2,
\eq
while when $\tilde{\Lambda}<0$, we find 
\bq
\Psi(L, t)=B_1\sin\left(\mu L +B_2\right),
\eq
here $A_1$, $A_2$, $B_1$ and $B_2$ are some parameters independent of L. Again none of these wavefunctions are normalizable with respect to the measure $L^{-1}dL$. 

 \subsubsection{$:\pi^2L:  \; = \hat{\pi}\hat{L}\hat{\pi}$}

In this case, we have 
\bq
\frac{\partial }{ \partial L}\left(L\frac{\partial \Psi}{\partial L}\right)-\epsilon_{\tilde{\Lambda}}\mu^2L\Psi=0.
\eq
When $\tilde{\Lambda}>0$, its general solution is given by the linear combination of modified Bessel functions of the first and second kind,  denoted, respectively,  by $I$ and $K$,   that is, 
\bq
\Psi(L,t)=C_3 I_0(\mu L)+C_4 K_0(\mu L).
\eq
 However, the normalizable condition with  the flat measure $dL$ in the interval $(0,+\infty)$ leads to 
\bq
C_3 = 0,\;\;\; C_4 = \frac{2}{\pi}\sqrt{\mu}.
\eq
 For $\tilde{\Lambda}=0$, we obtain
 \bq
 \Psi(L,t)=A_3\ln L+A_4,
 \eq
 which cannot be normalized in the interval $(0,+\infty)$.
When $\tilde{\Lambda}<0$, the general solution is given by 
\bq
\Psi(L,t)=B_3 J_0(\mu L)+B_4 Y_0(\mu L),
\eq
which is a linear combination of Bessel functions of the first and second kind. This wave function can't be normalized either.
 \subsubsection{$:\pi^2L:  \; = {\hat{\pi}}^2\hat{L}$}

In this case, we have 
\bq
\frac{\partial^2}{\partial L^2}\Big(L\Psi\Big)-\epsilon_{\tilde{\Lambda}}\mu^2L\Psi=0.
\eq
\lb{aa}
When $\tilde{\Lambda}>0$, the general solution of the above equation is given by, 
\bq
\Psi(L,t)=\frac{1}{L}\left(C_5 e^{- \mu L} + C_6  e^{ \mu L}\right),
\eq
where $C_5$ and $C_6$ are the integration constants. Similar to the first case, the wavefunction now is also not normalizable for any given $C_5$ and $C_6$
with respect to the  measure  $LdL$ in the interval $(0,+\infty)$. 
When $\tilde{\Lambda}=0$, the solution is 
\bq
\Psi(L,t)=A_5+\frac{A_6}{L},
\eq
For  $\tilde{\Lambda}<0$, we find
\bq
\Psi(L,t)=\frac{1}{L}\left[B_5\sin\left(\mu L+B_6\right)\right].
\eq
None of these wavefunctions are normalizable with respect to the measure $LdL$ in the interval $(0,+\infty)$.

\subsection{Simple Harmonic Oscillator}

In this subsection, we shall show that under canonical transformation the above system can be reduced to that of a simple harmonic oscillator.
By using the gauge freedom, we can always  set 
\bq
\lb{4.7}
N(t)=1, \;\;\; \;\;\; \;\;\;\; N_1=0.
\eq
Then,  applying the momentum constraint (\ref{4.20b}), the canonical Hamilton (\ref{4.19aa})
reduces to
\bq
\label {e}
H(L,\pi)=L \left[\frac{ \pi^2}{4\zeta^2(1-\lambda)}+2\zeta^2\Lambda\right],
\eq
with L given by Eq.(\ref{4.5}).
After the canonical transformation, 
\bqn
\lb{4.6a}
L&=&x^2,\;\;\; \pi=\frac{p}{2x},
\eqn
we find that Eq.(\ref{5.3bb}) yields $\{x, p\} =1$, and Eq.(\ref{e}) takes the form, 
\bq
\lb{4.9}
H' (x,p)=\frac{p^2}{16\zeta^2(1-\lambda)}+2\Lambda\zeta^2x^2.
\eq
However, this new Hamilton constraint (\ref{4.9}) can only be equivalent to the original one (\ref{e}) 
on the classical level. One can immediately understand this point when trying to find the solution to the corresponding  Wheeler-DeWitt equation,
\bq
H'(\hat{x},\hat{p})\Psi=0,
\eq
which yields no physical states due to non-vanishing of the energy of the ground state of the quantized oscillator. Hence, we employ 
the following ansatz for the quantum canonical transformation,
\bqn
\lb{4.9a}
\hat{L}&=&\hat{x}^2,\;\;\; \hat{\pi}=\frac{1}{4}\left(\frac{1}{\hat{x}}\hat{p}+\hat{p}\frac{1}{\hat{x}}\right).
\eqn
Correspondingly, some terms can be transformed into the  forms,
\bqn
\lb{4.9b}
\hat{L}\hat{\pi}^2&\Longrightarrow&\frac{\hat{p}^2}{4}+\frac{i}{2}\frac{1}{\hat{x}}\hat{p}-\frac{5}{16\hat{x}^2}, \\
\hat{\pi}\hat{L}\hat{\pi}&\Longrightarrow&\frac{\hat{p}^2}{4}-\frac{1}{16\hat{x}^2}, \\
\lb{4.9c}
\hat{\pi}^2\hat{L}&\Longrightarrow&\frac{\hat{p}^2}{4}-\frac{i}{2}\frac{1}{\hat{x}}\hat{p}+\frac{3}{16\hat{x}^2}.
\eqn
Now setting,
\bq
L\pi^2\ \mapsto \frac{1}{3}\left(\hat{L}\hat{\pi}^2+\hat{\pi}\hat{L}\hat{\pi}+\hat{\pi}^2\hat{L}\right),
\eq\
we find that the new Hamilton under the canonical transformation (\ref{4.9a}) is given by,
\bq
\tilde{H}=\frac{\hat{p}^2}{16\zeta^2(1-\lambda)}+2\Lambda\zeta^2\hat{x}^2-\frac{1}{64\zeta^2(1-\lambda)\hat{x}^2}.
\eq
Then, we  can introduce the creation and annihilation operators, 
\bqn
\lb{4.10}
a&=&c_0\left(x+\frac{ip}{8\zeta^2(1-\lambda)\sqrt{\tilde{\Lambda}}}\right),\nb\\
a^{\dagger}&=&c_0\left(x-\frac{ip}{8\zeta^2(1-\lambda)\sqrt{\tilde{\Lambda}}}\right),
\eqn
with $c_0 \equiv 2\zeta\sqrt{1-\lambda}\tilde{\Lambda}^{1/4}$, and
\bq
[a,a^\dagger]=1. 
\eq
In terms of $a$, $a^\dagger$ and $\hat{x}$, we find
\bq
\lb{Ha}
\tilde{H}= \hbar \omega \left(a^\dagger a +\frac{1}{2}\right)-\frac{1}{64\zeta^2(1-\lambda)\hat{x}^2}, 
\eq
where $\hbar \omega \equiv \sqrt{\tilde{\Lambda}}$. Clearly, to have a well defined vacuum, we must require $\tilde{\Lambda} > 0$, that is
\bq
\lb{4.17}
\frac{\Lambda}{1 - \lambda} > 0.
\eq
Then,   the Wheeler-DeWitt equation reads, 
\bq
\lb{4.18}
\tilde{H}(\hat{x},\hat{p})|\Psi\rangle=0.
\eq
Expanding $|\Psi \rangle$ in terms of the complete set $\{ |n\rangle \}$, 
\bq
\lb{4.19}
|\Psi\rangle=\sum\limits_{n=0}^{\infty}a_n  |n\rangle,
\eq
we find that 
\bqn
a_0+10\sqrt{2}a_2&=&0,\\
17a_1+14\sqrt{6}a_3&=&0,
\eqn 
and for $n\ge2$,  
\bqn
(4n-6)\sqrt{n(n-1)} a_{n-2}+(8n^2+8n+1)a_n\nb\\
+(4n+10)\sqrt{(n+1)(n+2)}a_{n+2}=0.
\eqn

Therefore, the wavefunction is given by
\bq
\lb{4.50}
\Psi(x) = \left<x|\Psi\right> = \sum_{n= 0}^{\infty}{a_n \psi_n(x)},
\eq
where $x = \sqrt{L}$, and 
\bqn
\lb{4.51}
 \psi_n(x) &\equiv& \left<x|n\right> =  \frac{(2\mu)^{2n+1}}{\pi^{1/4}\sqrt{2^nn!}}\nb\\
 && \times \left(x - \frac{1}{2\mu}\frac{d}{dx}\right)^n e^{-\mu x^2}.
 \eqn
Thus, we find that $\Psi(L) \propto e^{-\mu L}$, which is similar to the ones obtained by the Dirac quantization, although they are not precisely equal, as
we used two quite different approaches to obtain the corresponding  Hamiltons of quantum mechanics, as one can see from
Eqs.(\ref{5.3ba}) and (\ref{Ha}).  

From Eq.(\ref{4.10}), on the other hand, we find
\bqn
\lb{4.17a}
\big<m\left|L\right|n\big>&=& \ell_{HL}  \Big[(2n+1)\delta_{m,n} + \sqrt{n(n-1)}\delta_{m,n-2}\nb\\
&&  +\sqrt{(n+1)
(n+2)}\delta_{m,n+2}\Big],
\eqn
where $\ell_{HL} \equiv 1/(4c_0^2)$ denotes the meanvalue of the gauge-invariant length operator $L$ [cf. Eq.(\ref{4.5})] in the groundstate $\left|0\big>\right.$, i.e.,   
\bqn
\lb{4.17b}
\big<0\left|L\right|0\big> = \ell_{HL}.
\eqn

0

\subsection{Quantization of Spacetimes with $L(t) = 0$}

 It should be noted that, in the above studies,  either in terms of the Dirac quantization or in terms of the harmonic oscillator, we implicitly assumed $L(t) \not= 0$. Classically, this corresponds to the case
 studied in Sec. III.A, in which solutions exist only when $\tilde{\Lambda} > 0$, and the corresponding  space-time is de Sitter. But, quantum mechanically the quantization can be carried out for
 any given $\tilde{\Lambda} \in(- \infty, \infty)$.
 
  In addition, classical solutions exist even when $ L(t) =   \int_{-L_{\infty}}^{L_{\infty}}{\gamma(t, x) dx} =  0$.
 In   this case,   the classical solutions are given by  
 \bq
 \lb{a.a}
 \gamma(t,x)={\gamma}_0(x)\hat\gamma(t), 
 \eq
 where
 \bq
 \lb{a.aa}
 \int_{-L_{\infty}}^{L_{\infty}}{\gamma_0(x) dx} = 0,
 \eq
 that is, $\gamma_0(x)$ is an odd function of $x$. The function $\hat\gamma(t)$  satisfies the equation of motion,  Eq.(\ref{2.17}). 
 Of course, in this case the Hamiltonian  constraint (\ref{2.13a})  is satisfied identically, while the momentum constraint is satisfied, provided that
 \bq
 \lb{a.b}
 \pi(t, x) = \pi(t). 
 \eq
Then,  Eq.(\ref{2.17}) reads,  
 \bq
 \lb{a.c}
 2\ddot{\hat\gamma}\hat\gamma-\dot{\hat\gamma}^2+4\tilde{\Lambda}\hat\gamma^2=0,
 \eq
which  can be obtained from the effective action, 
\bq
\lb{a.d}
S_{\hat\gamma} = \int{{\cal{L}}_{\hat\gamma}dt},
\eq
where   
\bq
{\cal{L}}_{\hat\gamma} =\frac{\dot{\hat\gamma}^2}{\hat\gamma}-4\tilde{\Lambda}\hat\gamma.
\eq
After the Legendre transformation, the corresponding Hamilton turns out to be
\bq
H_{\hat\gamma}=\frac{1}{4} \hat\gamma {\pi_{\hat\gamma}}^2 +4\tilde{\Lambda}\hat\gamma,
\eq
where $\pi_{\hat\gamma}$ is  momentum conjugate of $\hat\gamma$. Remarkably, this Hamilton is nothing but the one  precisely given by  Eq.(\ref{5.3ba}).
Therefore,  its quantization can be carried out in the same
ways as we just did above: either by the standard Dirac quantization  or by the harmonic oscillator quantization. Thus, in the following we shall not repeat 
the above processes.

\section{Coupling with a scalar field}
\renewcommand{\theequation}{5.\arabic{equation}} \setcounter{equation}{0}

When the 2d HL gravity couples to a scalar field $\phi$, the total action becomes
\bq
\lb{5.1a}
S = S_{HL} + S_{\phi},
\eq
where $S_{\phi}$ denotes the action of the scalar field. To be power-counting renormalizable, the marginal terms of $S_{\phi}$ must be at least of dimension $2z$ with $z \ge d$.
Since  $\phi$ is dimensionless, one can see that  the marginal terms are $ \nabla_i\phi\nabla^i\phi$ and $a_i\nabla^i\phi$. Then, $S_{\phi}$  must take the form, 
\bqn
\lb{5.1b}
S_{\phi} &=& \int{dtdx N\sqrt{g}\Big[\frac{1}{2}\left(\partial_{\bot}\phi\right)^2  - \alpha_0 \left(\nabla_i\phi\right)^2 - V(\phi)}\nb\\
&&   - \alpha_1\phi \nabla^ia_i - \alpha_2 \phi a^i\nabla_i\phi\Big].
\eqn
Here  $\partial_{\bot} \equiv N^{-1} \left(\partial_t - N^i\nabla_i\right)$, $V(\phi)$ denotes the potential of the scalar field,  
and $\alpha_{n}$ are dimensionless coupling constants \footnote{Since the scalar field $\phi$ is dimensionless, 
these coefficients  in principle can be arbitrary functions of $\phi$. In this paper, 
we consider only the case where they are constants.}. In the relativistic limit, we have $\left(\alpha_0, \alpha_1, \alpha_2\right)_{GR} = (1, 0, 0)$.

\subsection{Classical Field Equations}

In the projectable case, we have $a_i = 0$ and the last two terms in Eq.(\ref{5.1b}) vanish. Then, the variations of the total action with respect to $N, \gamma, N_1$ and $\phi$, yield,
respectively, 
\bqn
\lb{5.2a}
&& \int{dx\Bigg\{\left[\frac{\dot{\gamma}^2}{\kappa\gamma}+8\zeta^2\Lambda\gamma\right] +\frac{2 c_{\phi}^2}{\gamma}\phi'^2}\nb \\
&& ~~~~~~~~~~~  + \left[2\gamma\dot{\phi}^2+4\gamma V\left(\phi\right)\right]\Bigg\}=0, \\
\lb{5.2b}
 &&\left(\frac{\dot{\gamma}}{\gamma}\right)^{.} + \frac{1}{2}\left(\frac{\dot{\gamma}}{\gamma}\right)^2+2\tilde\Lambda 
 = \kappa \left(\dot{\phi}^2+\frac{c_{\phi}^2}{\gamma^2}\phi'^2-2V\left(\phi\right)\right), \nb\\\\
\lb{5.2c}
&& \left(\frac{\dot{\gamma}}{\gamma}\right)' = 2\kappa\dot{\phi}\phi', \\
\lb{5.2d}
&& \left(\gamma \dot{\phi}\right)^{\cdot} -c_{\phi}^2 \left(\frac{\phi'}{\gamma}\right)' +\gamma\frac{dV\left(\phi\right)}{d\phi} = 0,
\eqn
where  $ c_{\phi}^2 \equiv 2\alpha_0$ must be non-negative in order for the scalar field to be stable,  and    
\bq
\kappa=\frac{1}{4\zeta^2\left(1-\lambda\right)}.
\eq
 
Note that in the vacuum case $\gamma$ is a function of $t$ only, as shown previously. However, because of the presence of the scalar field, now it in general  is a function of both $t$ and $x$. 
To compare it with the vacuum case, in the following let us consider the case $\gamma =\gamma_0(x) \gamma(t)$ only. In fact, as to be shown below, this is also the case where the corresponding 
 Hamiltonian constraint becomes local, while the momentum constraint can be solved explicitly. 
 
 Setting $\gamma = \gamma_0(x)\gamma(t)$, from Eq.(\ref{5.2c}) we can choose that $\phi= \phi(t)$. Then, Eqs.(\ref{5.2a}), (\ref{5.2b}) and (\ref{5.2d}) reduce, respectively, to
 \bqn
\lb{5.2aa}
&&\int{dx\left\{\frac{\dot{\gamma}^2}{\kappa\gamma}+8\zeta^2\Lambda\gamma +2\gamma\dot{\phi}^2+4\gamma V\left(\phi\right)\right\}}=0,\\
\lb{5.2bb}
 &&\left(\frac{\dot{\gamma}}{\gamma}\right)^{.} + \frac{1}{2}\left(\frac{\dot{\gamma}}{\gamma}\right)^2+2\tilde\Lambda 
 = \kappa \left(\dot{\phi}^2-2V\left(\phi\right)\right),  \\
\lb{5.2dd}
&&  \left(\gamma \dot{\phi}\right)^{\cdot}+\gamma\frac{dV\left(\phi\right)}{d\phi} =0.
\eqn

To solve the above equations, we further assume that   $V(\phi) = \tilde{\Lambda} = 0$. Then from Eq.(\ref{5.2dd}), we know 
\bq
\dot{\phi}=\frac{\phi_0}{\gamma(t)}, 
\eq
here $\phi_0$ is a constant. Combining with Eq.(\ref{5.2bb}), we derive an equation for $\gamma(t)$,
\bq
\lb{5.13aa}
\ddot{\gamma}(t)\gamma(t)-\frac{1}{2}\dot{\gamma}(t)^2=\kappa \phi_0^2.
\eq
One of the solutions can be easily obtained, and is given by
\bqn
\lb{5.13a}
\gamma(t)&=&\left(c_0+c_1 t\right)^2+\frac{\kappa \phi_0^2}{2c_1^2}, \\
\lb{5.13b}
\phi(t)&=&\sqrt{\frac{2}{\kappa}} \arctan\left(\sqrt{\frac{2}{\kappa}}\frac{c_1(c_0+c_1t)}{\phi_0}\right)+\phi_1, ~ ~~~
\eqn
where $c_0$, $c_1$ and $\phi_1$ are constants.
In order to make our solution consistent with the integral constraint (\ref{5.2aa}), we require $\gamma(t,x)$ to be an odd function of  $x$, so that
Eq.(\ref{2.16aa}) also holds here.  Keeping this in mind and then using the residual gauge freedom, we find the metric takes the form, 
\bq
\lb{5.14}
ds^2=-dt^2+\left(t^2+\epsilon_{\kappa} t_s^2 \right)^2dx^2,
\eq
here $\epsilon_{\kappa} \equiv {\mbox{sign}}(\kappa)$, and   
\bq
t_s^2 \equiv \frac{|\kappa| \phi_0^2}{2c_1^4}.
\eq

\begin{figure}[tbp]
\centering
\includegraphics[width=8cm]{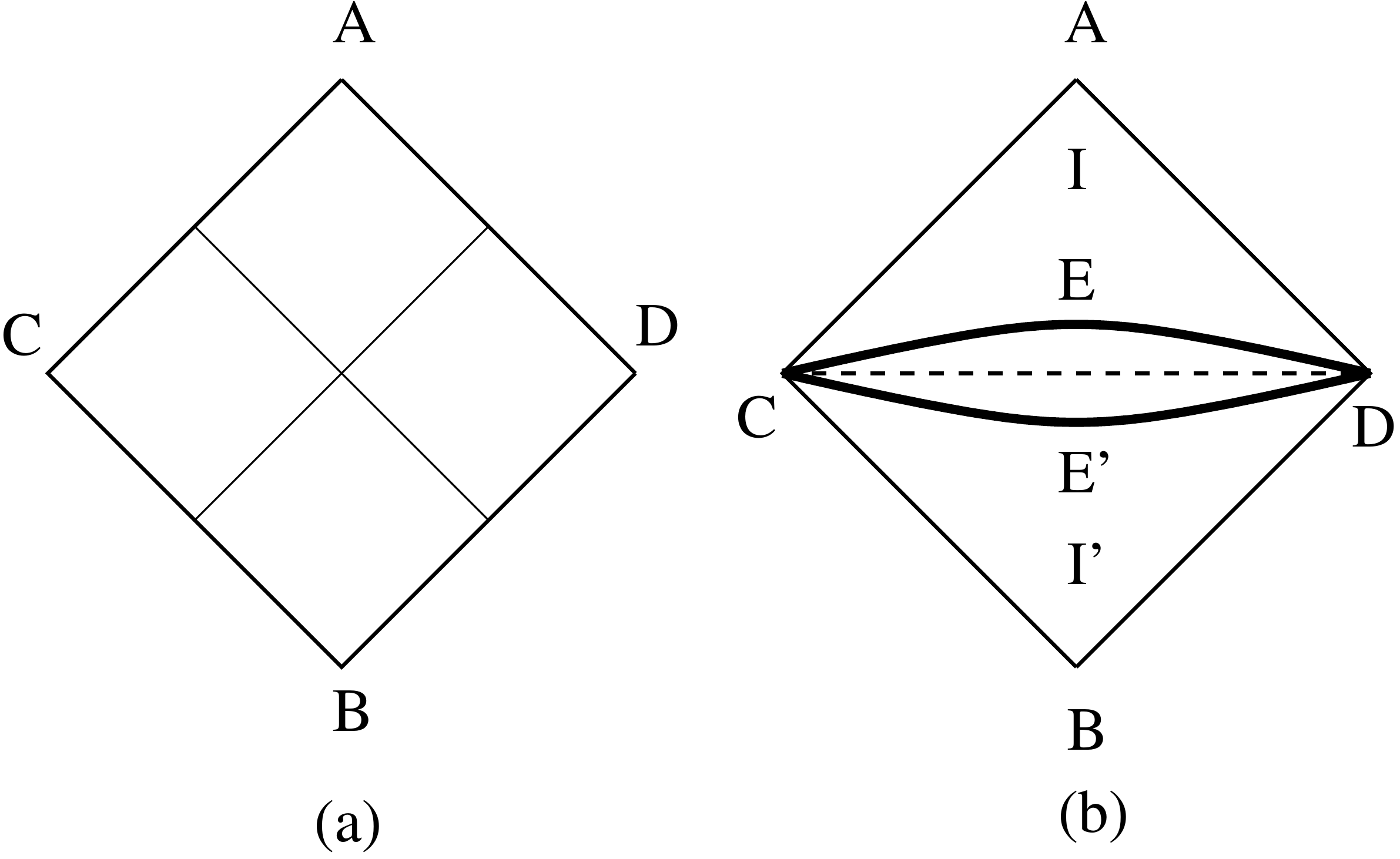}
\caption{ (a) The Penrose diagram for the solution (\ref{5.14}) with $\lambda \le 1$ (or $\kappa \ge 0$), in which the whole space-time is free of space-time singularities.
(b) The Penrose diagram for the solution  (\ref{5.14}) with $\lambda > 1$ (or $\kappa < 0$), in which the space-time is singular on $t = \pm t_s$, denoted by the thick solid curves 
$\widehat{CED}$ and $\widehat{CE'D}$. Thus, in this case the two regions $I$ and $I'$ are causally disconnected.}
 \label{fig4}
\end{figure}

Following what we did in Section III,  we can derive the extrinsic curvature $K$, Ricci scalar $R$, and the components of the tidal forces, given, respectively by, 
\bqn
K&=&-\frac{t}{2}R =-\frac{2c_1^2t}{t^2+\epsilon_{\kappa} t_s^2},\\
R_{(1)(1)}&=&-R_{(0)(0)}=\frac{2c_1^2}{t^2+\epsilon_{\kappa} t_s^2}.
\eqn
Therefore, the singularities of the spacetime are determined directly by the signs of $\kappa$. In particular,  
 if $\lambda \le 1$, the spacetime is free of space-time singularities.   For $\lambda >1$, on the other hand, there is a curvature singularity located at   
\bq
t=\pm t_s.  
\eq
The corresponding Penrose diagrams are given in Fig.\ref{fig4}.

It should be noted that, instead of imposing the condition (\ref{2.16aa}), we can set the integrand of the Hamiltonian constraint to zero. But, this will require   $C_1=0$, and the corresponding 
space-time is flat.

 \subsection{Hamiltonian Structure and Canonical Quantization}
 
When coupling with the scalar field, the Hamiltonian and momentum constraints become,
\bqn
\lb{5.7}
&& \int{dx \mathcal{H}(x)}= \int{dx \left[\frac{\pi^2\gamma}{4\zeta^2(1-\lambda)}+2\Lambda \zeta^2 \gamma +\frac{\pi_{\phi}^2}{2\gamma}\right.}\nb\\
&& ~~~~~~~~~~~~~~~~~~~  \left. +\frac{\alpha_0\phi'^2}{\gamma}+\gamma V(\phi)\right], \\
\lb{5.8}
&& \mathcal{H}_1 = -\frac{\pi'}{\gamma}+\frac{\pi_{\phi}\phi'}{\gamma^2},
\eqn
here $\pi_{\phi}$ denotes the canonical moment conjugate to the scalar field $\phi$. Similarly, the Poisson brackets of the two constraints are given by, 
 \bqn
\{\mathcal{H}(x),\mathcal{H}_1(x')\}&=&\frac{\mathcal{H}(x')\delta_x(x-x')}{\gamma^2(x')}+\frac{\pi\mathcal{H}_1\delta(x-x')}{\zeta^2(1-\lambda)}, \nb\\
\{\mathcal{H}_1(x),\mathcal{H}_1(x')\}&=&\frac{2\mathcal{H}_1(x)\delta_{x'}(x-x')}{\gamma^2(x)}\nb\\
&& +\frac{2\gamma'\mathcal{H}_1}{\gamma^3}\delta(x-x') - \frac{\mathcal{H}'_1}{\gamma^2}\delta(x-x').\nb\\
\eqn
For the non-local Hamiltonian constraint we also find
\bq
\left\{\int{dx\mathcal{H}(x)},\int{dx'\mathcal{H}(x')}\right\}=0,
\eq
as long as $\pi_{\phi}\phi'/\gamma^2$ vanishes on boundaries. 

In the rest of this paper, we only consider the quantization of   the system for the  case,
\bq
\phi'=0 = \pi',
\eq
in order to compare with what we obtained in the pure gravity case.  As a matter of fact, this also makes the problem considerably simpler and   
tractable. 
Under the above  assumption, the Hamiltonian constraint reads, 
\bq
\lb{5.8a}
H(t)=\frac{\pi^2 L}{4\zeta^2(1-\lambda)}+2\Lambda \zeta^2 L +\frac{L{\dot{\phi}}^2}{2}+L V(\phi) \simeq 0.
\eq
It must be noted  that in writing down the above expression, we performed the spatial integration and used the fact that 
\bq
\lb{5.8c}
\pi_{\phi}=\gamma\dot{\phi},
\eq
with  the gauge choice $N=1$ and $N_1=0$. 
On the other hand,  from the canonical relation,  
\bq
\{\phi(x),\pi_{\phi}(y)\}=\delta(x-y),
\eq
we can integrating both sides with respect to the spatial coordinates x and y, and then use Eq.(\ref{5.8c}) and the constraint $\phi=\phi(t)$, to obtain
\bq
\{\phi(t), L(t)\dot{\phi}(t)\}=1,
\eq
which enables us to identify $\pi_{\phi}$ as $\pi_{\phi} = L\dot{\phi}$. Now making this substitution in the Hamiltonian constraint (\ref{5.8a}),
 we find the Hamilton with two discrete physical degrees of freedom, $L$ and $\phi$, takes the form, 
\bq
\lb{5.8b}
H(t)=\frac{\pi^2 L}{4\zeta^2(1-\lambda)}+2\Lambda \zeta^2 L +\frac{\pi_{\phi}^2}{2L}+L V(\phi).
\eq
Thus, the Wheeler-Dewitt equation now reads, 
\bq
\lb{5.9}
\hat{H}(t)\Psi(L,\phi; t)=0.
\eq
If we further assume that the potential of the scalar field can be  ignored, $V(\phi) \simeq 0$, we are able to find solutions to Eq.(\ref{5.9}) by separation of variables.
In this case, assuming 
\bq
\lb{5.9a}
\Psi(L,\phi)=X(L)Y(\phi),
\eq
we obtain two independent equations, 
\bqn
\lb{5.9b} 
Y''(\phi)+mY(\phi)=0,\\
\lb{5.9c} 
\left[L\pi^2\right]X(L)+\left(\mu^2L+\epsilon_{\lambda}\frac{m \mu}{2\sqrt{\tilde{\Lambda}}L}\right)X(L)=0.
\eqn
Here $[L\pi^2]$ means some specific ordering of L and $\pi$, $m$ is an undetermined parameter,  $\mu$ is given as in the pure gravity case, and
$\epsilon_{\lambda}$ is   one for $\lambda<1$ and negative one for $\lambda>1$. Eq.(\ref{5.9b}) has the general solution,
\bq
\lb{5.9d}
Y(\phi) = D_1 \sin\left(\sqrt{m}\phi + D_2\right),  
\eq
where $D_{1,2}$ are tow integration and possibly complex  constants. To solve Eq.(\ref{5.9c}), just as in the pure gravity case, there are   three different orderings, which will be considered
below, separately. 

 \subsubsection{$:\pi^2L:  \; = \hat{L}\hat{\pi}^2$}
 
In this case, the Hamiltonian constraint reads
\bq
L^2X''-\left(\epsilon_{\tilde{\Lambda}}\mu^2L^2+k^2\right)X=0,
\eq
where $k^2\equiv 2\epsilon_{\lambda}m\zeta^2|1-\lambda|$ and $\epsilon_{\tilde{\Lambda}}$ is defined in Sec. IV. For $\tilde{\Lambda}>0$,
the general solution is given by the linear combination of modified Bessel functions of the first and second kind, denoted by $I_\nu$ and $K_\nu$, respectively, that is,
\bq 
X =\sqrt{L}\left\{C_1I_\nu(L\mu)+C_2K_\nu(L\mu)\right\},
\eq
Here $\nu\equiv \sqrt{1+4k^2}/2$. Generally, this wave-function is not normalizable with respect to 
the measure $dL/L$ in the interval $(0,+\infty)$. However, if $|\Re(\nu)|<1/2$, we have the normalized function, given by
\bq
X_{\text{norm}}=\frac{1}{\pi}\sqrt{\frac{4\mu}{\sec\left(\pi\nu\right)}}K_\nu\left(L\mu\right).
\eq
In this particular case for $-1/4\leq k^2<0$, depending on the value of $\lambda$, the parameter $m$ can be either positive or negative. When it is positive,   in order to have a 
normalizable wave function $\Psi(L, \phi)$ of Eq.(\ref{5.9a}), we need to restrict  the domain of  $\phi$ to some finite region, for example $(0, 2\pi)$, then it would be 
straightforward to normalize $Y(\phi)$ from Eq.(\ref{5.9d}) in that finite region. When $m$ is negative, $Y(\phi)$ can be normalizable either in the region $\phi \in (-\infty, 0)$ or
$\phi \in (0, \infty)$, so can the wavefunction $\Psi(L, \phi)$. 

For $\tilde{\Lambda}=0$, the solution is given by
\bq
X=\sqrt{L}\left(A_1L^{+\nu}+A_2L^{-\nu}\right),
\eq
while for  $\tilde{\Lambda}<0$, we find
\bq
X=\sqrt{L}\left(B_1J_{\nu}(\mu L)+B_2Y_{\nu}(\mu L)\right).
\eq
Here $\nu$ is defined as in the case $\tilde{\Lambda}>0$. None of these two wave functions are normalizable with respect to the measure $L^{-1}dL$ in the interval $(0,+\infty)$.

\subsubsection{$:\pi^2L:  \; = \hat{\pi}\hat{L}\hat{\pi}$} 
 
In this case, we have 
\bq
L^2X''+LX'-\left(\epsilon_{\tilde{\Lambda}}\mu^2L^2+k^2\right)X=0.
\eq
Thus, for $\tilde{\Lambda}>0$, the general solution is given by, 
\bq
X=C_1I_k(L\mu)+C_2K_k(L\mu).
\eq
Again, for $0\leq k^2<1/4$, we have the normalized function $X(L)$ given by
\bq
X_{\text{norm}}=\frac{1}{\pi}\sqrt{\frac{4\mu}{\sec\left(\pi k\right)}}K_k\left(L\mu\right).
\eq
Similar to the last case,  $Y(\phi)$ is normalized only in some restricted domains, depending on the signs of $m$. 

When $\tilde{\Lambda}=0$, its general solution is 
\bq
X=A_1L^k+A_2L^{-k},
\eq
while for $\tilde{\Lambda}<0$, it is given by
\bq
X=B_1 J_k(\mu L)+ B_2 Y_k(\mu L).
\eq
It can be shown that none of these two wavefunctions are normalizable in the interval $(0,+\infty)$. 

\subsubsection{$:\pi^2L:  \; = \hat{\pi}^2\hat{L}$} 
 
In this case, we have 
\bq
L^2X''+2LX'-\left(\epsilon_{\tilde{\Lambda}}\mu^2L^2+k^2\right)X=0.
\eq
Then, for $\tilde{\Lambda}>0$, we find
\bq
X=C_1j_{-\nu-1/2}(-i L\mu)+C_2 y_{-\nu-1/2}(- i L\mu),
\eq
here $j_{\nu}$, $y_{\nu}$ denote spherical Bessel functions of the first and second kind. 
When $\tilde{\Lambda}=0$, we find that
\bq
X=L^{-1/2}\left(A_1 L^{\nu}+A_2 L^{-\nu}\right),
\eq
while for $\tilde{\Lambda}<0$, we have
\bq
X=C_1j_{\nu-1/2}(\mu L)+C_2 y_{\nu-1/2}(\mu L).
\eq
It can be shown that in this case none of these wave functions are normalizable with respect to the measure $LdL$ in the interval $(0,+\infty)$.

\subsection{Two Interacting Simple Harmonic Oscillators}  

Similar to what we have done in the pure gravity case, we can also treat  the Hamilton given by Eq.(\ref{5.8b}) as consisting of harmonic oscillators. 
 To this goal, let us first make  the transformations
\bqn
\lb{5.32}
L(t) &=&y_1^2(t) -y_2^2(t),\nb\\
\phi(t) &=& \sqrt{2\zeta^2(\lambda - 1)}\ln\left(\frac{y_1(t) +y_2(t)}{y_1(t) -y_2(t)}\right),
\eqn
for which we are able to convert Eq.(\ref{5.8c}) into the form, 
\bqn
\lb{5.33}
{\mathcal{L}}&=& \frac{1}{2}m\Big[\left(\dot{y}_1^2 - \omega^2y_1^2\right)  - \left(\dot{y}_2^2 - \omega^2y_2^2\right)\Big]\nb\\
&& - V_{e}(y_1, y_2),
\eqn
but now with
\bqn
\lb{5.34}
&& m \equiv 8(1-\lambda)\zeta^2,\;\;\;
\omega^2 \equiv \frac{\Lambda}{2(1-\lambda)},\nb\\
&& V_{e}(y_1, y_2) \equiv \left(y_1^2-y_2^2\right) V\Big(\phi(y_1,y_2)\Big).
\eqn

Clearly, Eq.(\ref{5.33}) describes the interaction between two simple harmonic oscillators, one with positive energy and the other with negative energy. Thus, in order for
the system  to have a total positive energy, the interaction between them is important. 

To process further, we need to consider particular potential $V(\phi)$, which will be model-dependent. So, in the following we shall not pursue the quantization alone this direction 
further. 

\subsection{Quantization of Spacetimes with $L(t) = 0$}

Just like in the pure gravity part, when $L(t)=0$, we  again have $\gamma(t,x)={\gamma}_0(x)\hat\gamma(t)$,
where  ${\gamma}_0(x)$ is an odd function of $x$, so Eq.(\ref{a.aa}) is satisfied, which in turn guarantees that the Hamiltonian constraint   (\ref{5.7}) 
is   automatically satisfied, while the momentum constraint (\ref{5.8}) will be also satisfied when $\pi = \pi(t)$ and ${\phi} = {\phi}(t)$.
Then,  the equations of motion (\ref{5.2b}) and (\ref{5.2d}) can be obtained from the effective   Lagrange, 
\bq
{\cal{L}}_{(\gamma,\phi)} \equiv \frac{\dot{\gamma}^2}{\gamma}+2\kappa \gamma \dot{\phi}^2-4 \tilde{\Lambda}\gamma-4\kappa \gamma V(\phi).
\eq
Note that in writing the above equation, we had dropped the hat from $\gamma$. Then, the corresponding Hamilton is given by 
\bq
H_{(\gamma,\phi)} =\frac{\pi_{\gamma}^2}{4} \gamma+\frac{\pi_{\phi}^2}{8\kappa\gamma}+4\tilde{\Lambda}\gamma+4\kappa\gamma V(\phi),
\eq
which has the same form as the Hamilton given by Eq.(\ref{5.8b}). Therefore, its  quantization can be followed precisely what we did in the above, which will not be repeated
here.

 \section{Conclusions}
 \renewcommand{\theequation}{6.\arabic{equation}} \setcounter{equation}{0}

In this paper, we have studied the quantization of the (1+1)-dimensional projectable HL gravity. In particular, after giving a brief review of the
theory with or without the projectability condition  in Sec. II, we have devoted Sec. III to study vacuum solution of the classical HL gravity, and found all 
the solutions in the projectable case. These solutions can be divided into several classes, and each of them have different local and global properties. 
Their corresponding Penrose diagrams are given, respectively, by Figs. \ref{fig1},  \ref{fig2} and \ref{fig3}. 

In Sec. IV, after  working out the Hamiltonian structure and solving the momentum constraint explicitly for the projectable  vacuum HL gravity, we have showed 
that the resulting Hamilton can be quantized  by following  the standard Dirac quantization. When moving from the classical Hamilton to the quantum mechanical one,
ordering ambiguity always appears. We have found that for some orderings the corresponding wavefunctions are normalizable. In addition, the  Hamilton can also be 
written in the form of a simple harmonic oscillator, whereby its quantization can be carried out in the standard way. Again, the orderings of relevant operators play
an essential role, so that the Weeler-DeWitt equation $\hat{H}\left|\Psi\right> = 0$ has non-trivial solutions.
%

In Sec. V, we have extended the studies carried out in Sec. IV to couple  minimally with a scalar field, and solved  the momentum constraint
 in the case where the fundamental variables are functions of time only. In this particular case, the quantization of the coupled system can also be carried out
by the standard Dirac process. After writing the corresponding Hamilton in terms of two interacting harmonic oscillators, we have found  that 
one of them has positive energy, and  the other has negative energy, once the interaction is turned off.

A remarkable feature is that the space-time can be quantized, even it classically has various singularities [cf. Figs.\ref{fig1} and \ref{fig4}]. In this sense, the classical singularities are 
indeed smoothed out by the quantum effects.

It should be noted that in this paper we have mainly studied the case with the projectability condition. It would be very interesting to see what will happen if such condition is relaxed. 
We wish to come back to this case soon.

 \section*{Acknowledgements}

 We thank B. Shakerin for the participation  of this project in its early stage. Our thanks  also go to Dr. Tao Zhu for valuable discussions and suggestions.
 This work is supported in part by DOE, DE-FG02-10ER41692 (A.W.),
Ci\^encia Sem Fronteiras, No. 004/2013 - DRI/CAPES, Brazil (A.W.), and Natural Science Foundation of
China (NSFC) Grant  No. 11375153 (A.W.). B.F.L. would  like to thank Baylor University for 
support through the Baylor  graduate fellowship.

\end{document}